\hfuzz 1pt
\font\titlefont=cmbx10 scaled\magstep1
\magnification=\magstep1

\null
\vskip 1.5cm
\centerline{\titlefont COMPLETE POSITIVITY}
\smallskip
\centerline{\titlefont AND THE K-$\overline{\hbox{K}}$ SYSTEM}
\vskip 2.5cm
\centerline{\bf F. Benatti}
\smallskip
\centerline{Dipartimento di Fisica Teorica, Universit\`a di Trieste}
\centerline{Strada Costiera 11, 34014 Trieste, Italy}
\centerline{and}
\centerline{Istituto Nazionale di Fisica Nucleare, Sezione di 
Trieste}
\vskip 1cm
\centerline{\bf R. Floreanini}
\smallskip
\centerline{Istituto Nazionale di Fisica Nucleare, Sezione di 
Trieste}
\centerline{Dipartimento di Fisica Teorica, Universit\`a di Trieste}
\centerline{Strada Costiera 11, 34014 Trieste, Italy}
\vskip 2cm
\centerline{\bf Abstract}
\smallskip
\midinsert
\narrower\narrower\noindent
Models that provide experimentally testable violations of ordinary
Quantum Mechanics have been recently proposed.
These models are based on non-unitary time evolutions 
of density matrices that are generated by linear positive maps.
We discuss the consequences of imposing a stronger condition on those
maps, known as complete positivity. 
It turns out that experimental data on the neutral kaon system
giving upper bounds to the 
parameters characterizing positive maps, also give bounds to those
determining completely positive ones.
\endinsert
\bigskip
\vfil\eject

Since its discovery, the $K$-$\overline K$ system has been used as a testing
ground for fundamental questions in Quantum Mechanics. 
At first, experiments with kaons gave precise evidence to violations of 
discrete symmetries.
Nowadays, it seems possible to test the completeness of Quantum Mechanics
itself, particularly thanks to the high luminosity of the new particle
factories being built [1].

The physical intuition underlying all the recently proposed models that give
rise to violations of Quantum Mechanics stems back to the idea that
the ``foam'' structure of space-time at Planck's
length produces loss of phase-coherence [2, 3].
Therefore, mixed states (density matrices) and no more pure states
are fundamental for the description of physical systems, and accordingly
time-evolutions transforming pure states into mixed ones are allowed.

Quantum evolutions based on the standard Liouville equation
$$
{\partial\rho(t)\over\partial t}=-i[H,\rho(t)]\ , \eqno(1)
$$ 
always preserve purity of states; indeed, if the initial density matrix 
$\rho(0)$
is pure, {\it i.e.} $[\rho(0)]^2=\rho(0)$, then for all times:
$$
[\rho(t)]^2=\rho(t)\ . \eqno(2)
$$
When the Hamiltonian $H$ is not hermitian, Eq.(1) becomes
$$
{\partial\rho(t)\over\partial t}=-iH\, \rho(t)+i\rho(t)\, H^\dagger\ . \eqno(3)
$$ 
This happens in the phenomenological description
of the decay properties of the 
$K$-$\overline K$ system (Weisskopf-Wigner approximation). 
We notice that Eq.(3) does not yield condition (2),
since $H\neq H^\dagger$ entails loss of probability. 
However, this is not loss of coherence: pure states remain pure,
except for a (time-dependent) normalization. In fact, while (1) generates
a time-evolution with group property, (3) produces a decay-like evolution,
hence the group composition property holds either for positive or negative 
times.

Loss of coherence shows up when further terms are added to the r.h.s. of (3).
We will consider minimal modifications of the time evolution equation
of the type
$$
{\partial\rho(t)\over\partial t}=-iH\, \rho(t)+i\rho(t)\, H^\dagger+h[\rho(t)]
\ , \eqno(4)
$$
in such a way that the additional piece $h[\rho]$ generates an evolution map
$\tau_t: \rho(0)\mapsto\rho(t)$ that is linear
and transforms density matrices into density matrices. 
This means that probability is preserved by $\tau_t$:
${\rm Tr}(\rho(t))={\rm Tr}(\rho(0))$, and that loss of probability
might only come from non-hermitian parts of the Hamiltonian.
Further, density matrices are positive operators;
this means that any density matrix $\rho$ can be written as 
$\rho=\sigma^\dagger\sigma$, for a suitable operator $\sigma$.
Asking for $\rho(t)$ to be again a density matrix implies that
$\tau_t$ has to be positive as a linear transformation.
More generally, these minimal requests follow from the assumption
of the existence of a scattering operator mapping linearly in-states
into out-states [3].

A stronger request to be made on the map $h[\rho]$ 
is that the evolution $\tau_t$ it generates,
be not only positive, but
``completely positive'' [4-7]. This requirement has a precise
physical meaning. Suppose we have two separate systems $A$ and $B$,
described by the density matrices $\rho_A$ and $\rho_B$. If $\rho_A$
evolves into a new state by means of the dynamics of $A$ only, then 
$B$ can not be affected by that changement, and $\rho_B$ remains
unaltered. In other words, two physically uncorrelated systems
can not influence each other [7].

More precisely, suppose the evolution in $A$ transforms the state
$\rho_A$ into $\omega\big[\rho_A\big]$; then, complete positivity
amounts to requiring that the map defined by
$$
\tilde{\omega}\big[\rho_A\otimes\rho_B\big]=\omega\big[\rho_A\big]
\otimes\rho_B\ , \eqno(5)
$$
be positive for any separate system $B$.

The simplest physical instance of a completely positive evolution is
any unitary one, {\it i.e.} those generated by Eq.(1). More interestingly, 
think of the two separate systems $A$ and $B$ as forming a global
system evolving under a unitary group $U(t)$.
If $\rho$ is the state of the compound system, the reduced states of
the subsystems are given by the partial traces over the unwanted degrees of
freedom; with obvious notations: $\rho_A={\rm Tr}_{B}(\rho)$, 
$\rho_{B}={\rm Tr}_{A}(\rho)$.
Let $O_A$ be an observable of the subsystem $A$.
As an observable of the global system, it can be written as
$O=O_A\otimes{\bf 1}_{B}$.
Its mean value at time $t$ follows from:
$$
\langle O(t)\rangle_{\rho}={\rm Tr}\left(\rho\, U(t) O U^\dagger(t) \right)=
{\rm Tr}_A\left({\rm Tr}_{B}\left[U^\dagger(t)\rho U(t)\right] O_A\right)\ .
\eqno(6)
$$
One can check that the map 
$\omega_t:\rho_A\mapsto\rho_A(t)\equiv
{\rm Tr}_{B}\left[U^\dagger(t)\rho U(t)\right]$ 
is completely positive and maps density matrices into density
matrices, ${\rm Tr}_A(\rho_A(t))={\rm Tr}_A(\rho_A)$ [4].

Notice however, that the sub-dynamics $\omega_t$ on $A$ does not 
in general satisfy a group composition law 
$\omega_t\circ\omega_s=\omega_{t+s}$, not even
for $t$, $s$ positive (semigroup composition law).
On the other hand, an obvious physical requirement for the evolution
$\tau_t$ generated by $h[\rho]$ in (4) is that it forms a semigroup:
$$
\tau_t\circ\tau_s=\tau_{t+s}\ ,\quad t,\ s\geq0\ .\eqno(7)
$$
The two requests, semigroup composition law and complete
positivity, fixes the form of the generator of $\tau_t$ [5, 6],
and the linear map $h[\rho(t)]$ in Eq.(4) must read:
$$
h[\rho(t)]=
-{1\over2}\Big(\sum_j A_j^\dagger A_j\, \rho(t)+\rho(t)\sum_j A_j^\dagger A_j
\Big)+\sum_j A_j\rho(t) A_j^\dagger\ ,\eqno(8)
$$
where the $A_j$ are suitable operators, such that 
$R\equiv\sum_j A_j^\dagger A_j$ be bounded. 

The map $h$ is further constrained by the request that 
the (von Neumann) entropy, $S[\rho]=-{\rm Tr}(\rho\ln\rho)$ increases:
$$
{d S[\rho(t)]\over dt}\geq 0\ . \eqno(9)
$$
This corresponds to the idea that information can be lost, for example
due to quantum gravity effects. From Eq.(8), this
is most simply achieved by asking that $A_j^\dagger=A_j$.[8] 

Some remarks are in order at this point. Suppose that the system $B$
considered above is a heat bath. Then, there exists a canonical
technique to produce a completely positive semigroup driving the subsystem $A$ 
(a so called quantum dynamical semigroup [4-6]).
It is based on two approximations: first one 
assumes a weak interaction between $A$ and the
reservoir $B$, and second that the characteristic times of $B$ be
much shorter than those of $A$. Technically, one performs
a weak coupling limit (or van Hove limit) on Eq.(6) and ends up with a
generator of the form (8) for the dynamics of $A$.

In this respect, let us point out that there are two philosophically
quite different attitudes in dealing with the evolution generated
by Eq.(4) and Eq.(8): either one considers it as an effective description
derived from a more fundamental dynamics,
ore one deals with that evolution as fundamental in itself.
An example of the latter point of view is discussed in [9], where the origin
of the additional piece $h[\rho]$ in (4) is traced back to the ``foamy''
structure of space-time, which is intrinsically unobservable,
except for its effects on loss of quantum coherence.
As a consequence, the physical time-evolution violates ordinary
Quantum Mechanics.

The situation of a subsystem weakly coupled to a heat bath discussed before
is instead an example of the former point of view. In this case,
Quantum Mechanics is not violated at the level of the global dynamics of the
subsystem plus reservoir, but only by the reduced dynamics.
Further, notice that quantum dynamical semigroups generated by (8)
arise as effective dynamics also in situations involving
environments quite different from heat baths: for instance, from the 
interaction
of a microsystem with a macroscopic measuring apparatus [10].
Therefore, treating the modified evolution as an effective dynamics
allows discussing in principle more general physical conditions.

As a final remark, we stress that, in general, Eq.(4) and Eq.(8) do not
respect conservation laws even in presence of symmetries [9].
While from a phenomenological point of view this fact can be tolerated since
in any case symmetries correspond to quantities that are conserved  
by the global dynamics,  this is no longer true if Eq.(4) and Eq.(8)
are regarded as fundamental. The consequences of this fact 
have been discussed in [11, 12].
In any case, notice that these violations can be confined within space-time
regions of dimensions of the order of Planck's length [13].

In what follows we take on the ``phenomenological'' attitude
and consider Eq.(4) and Eq.(8) as effective evolution equations and apply
them to study the $K$-$\overline K$ system. The mixing of these two 
particles can be modeled by means of a two-dimensional Hilbert space.
We use the $CP$-eigenstates $|K_1\rangle$ and $|K_2\rangle$ as an orthonormal
basis:
$$
|K_1\rangle={1\over\sqrt{2}}\Big[|K\rangle+|\overline{K}\rangle\Big]\ ,\quad
|K_2\rangle={1\over\sqrt{2}}\Big[|K\rangle-|\overline{K}\rangle\Big]\ .
\eqno(10)
$$
With respect to this basis a density matrix will be written as:
$$
\rho=\left(\matrix{
\rho_{11}&\rho_{12}\cr
\rho^*_{12}&\rho_{22}}\right)\ , \eqno(11)
$$
where $*$ signifies complex conjugation.
In the standard quantum mechanical description, 
the decay properties of the $K$-$\overline K$ system are very well
captured by the Weisskopf-Wigner Hamiltonian which is
conventionally written as $H=M-{i\over 2}{\mit\Gamma}$, 
where $M$ and $\mit\Gamma$
are hermitian $2\times 2$ matrices.
We use the convention of [14] and characterize them in terms of the 
complex parameters $\epsilon_S$, $\epsilon_L$, appearing in the eigenstates, 
$$
|K_S\rangle=N_S\left(\matrix{1\cr\epsilon_S}\right)\ ,\quad
|K_L\rangle=N_L\left(\matrix{\epsilon_L\cr 1}\right)\ ,
\eqno(12)
$$
and the four real parameters, $m_S$, $\gamma_S$ and $m_L$, $\gamma_L$
characterizing the eigenvalues of $H$: 
$$
\lambda_S=m_S-{i\over 2}\gamma_S\ ,\quad
\lambda_L=m_L-{i\over 2}\gamma_L\ ;\eqno(13)
$$
$N_S$ and $N_L$ are normalization factors.

Motivated by the previous considerations about complete positivity
and entropy increase, we will
generalize the Weisskopf-Wigner evolution equation (3) for
the $K$-$\overline K$ system and write: 
$$
{\partial\rho(t)\over\partial t}=-iH\, \rho(t)+i\rho(t)\, H^\dagger 
-{1\over 2}\Big[R\, \rho(t) + \rho(t)\, R\Big]+\sum_j A_j\rho(t) A_j\ ,
\eqno(14)
$$
where the $A_j$ are now hermitian $2\times 2$ matrices.

Both $\rho$ and the $A_j$'s can be expanded in terms of Pauli
matrices $\sigma_i$ and the identity $\sigma_0$:
$$
\rho=\rho_\alpha\sigma_\alpha\ ,\quad A_j=a^j_\alpha\sigma_\alpha\ ,
\qquad\alpha=0,1,2,3\ .
\eqno(15)
$$
For notational convenience, we introduce the following three vectors 
with components labeled by the index $j$
$$
\vec{a}_1=(a^j_1)\ ,\quad \vec{a}_2=(a^j_2)\ , \quad
\vec{a}_3=(a^j_3)\ .\eqno(16)
$$
In this way one can easily work out the action of
$$
h[\rho]=-{1\over 2}\Big[R\, \rho(t) + \rho(t)\, R\Big]+\sum_j A_j\rho(t) A_j\ ,
\eqno(17)
$$
as a $4\times 4$ matrix $[h_{\alpha\beta}]$ 
on the column vector with components $(\rho_0,\rho_1,\rho_2,\rho_3)$;
the entries $h_{\alpha\beta}$
can be expressed in terms of the scalar products of the vectors (16):
$$
h_{0\beta}=h_{\alpha0}=0\ ,\quad h_{rs}=2\,\vec{a}_r\cdot\vec{a}_s
-2\, \delta_{rs}\, \sum_{k=1}^3\vec{a}_k\cdot\vec{a}_k\ ,\quad
r,s=1,2,3\ .
\eqno(18)
$$
Notice that the $a^j_0$ components of the matrices $A_j$ do not appear
and the number of free parameters is six: the lengths of the vectors
$\vec{a}_1$, $\vec{a}_2$, $\vec{a}_3$, and the corresponding angles
between them. Therefore, there is a minimal choice for the matrices
$A_j$, the one for which: $j=1,2,3$, the components $a^j_0=\,0$, and
$\vec a_1$, $\vec a_2$ and $\vec a_3$ are linearly independent
three-dimensional real vectors.

From Eq.(18), it follows that the three-dimensional matrix
$[h_{rs}]$ is symmetric, and further that its entries satisfy the following
inequalities:
$$
\eqalign{&h_{11}\geq h_{22}+h_{33}\ ,\phantom{\big(h_{22}\big)^2}\cr
         &h_{22}\geq h_{11}+h_{33}\ ,\phantom{\big(h_{22}\big)^2}\cr
         &h_{33}\geq h_{11}+h_{22}\ ,\phantom{\big(h_{22}\big)^2}\cr}\quad
\eqalign{&h_{12}^2\leq h_{33}^2-\big(h_{11}-h_{22}\big)^2\ ,\cr
         &h_{13}^2\leq h_{22}^2-\big(h_{11}-h_{33}\big)^2\ ,\cr
         &h_{23}^2\leq h_{11}^2-\big(h_{22}-h_{33}\big)^2\ .\cr}
\eqno(19)
$$
These are direct consequence of the positivity of the norms and
of the Schwartz inequalities for the scalar products of the three vectors 
$\vec a_1$, $\vec a_2$ and $\vec a_3$.

Note that if one of the diagonal entries of $[h_{rs}]$ is zero,
then the other two must be equal and the off-diagonal terms
must vanish. For example, taking $h_{11}=\, 0$, the four-dimensional
matrix $[h_{\alpha\beta}]$ reduces to:
$$
[h_{\alpha\beta}]=\left(\matrix{0&0&0&0\cr
                              0&0&0&0\cr
                              0&0&h_{22}&0\cr
                              0&0&0&h_{22}\cr}\right)\ .\eqno(20)
$$

Comparing this result with the quantum mechanical violating term $h[\rho(t)]$
proposed in [9], we see that there the evolution equation for the
$K$-$\overline K$ system gives rise to a completely positive semigroup
only if $h[\rho]$ acts as in (20); in the notation of [9], this means 
that the three phenomenological parameters must fulfill:
$\alpha=\gamma$ and $\beta=\,0$.

The choice made in [9] for the map $h[\rho]$ is a very particular one;
in fact, despite our requirement of complete positivity, we have three more 
phenomenological parameters, although subjected to the conditions (19).
The reason for this is that in [9] it is argued that modifications
of Quantum Mechanics coming from quantum gravity might violate
strangeness conservation only as a second order effect. 
In our approach, such an attitude would force us to
take $h_{11}=\,0$ in (18), and thus to reduce to (20).

We would like to point out 
that violation of the quantum mechanical time-evolution
might be due to other effects than gravitational
ones. These other effects, as briefly mentioned above,
could be accounted for by the request of complete positivity.
The larger number of parameters in $[h_{\alpha\beta}]$ thus available
need to be fixed from the experimental data. We find it remarkable
that the condition of complete positivity might be tested experimentally.

In the remaining part of this letter we shall obtain
some order of magnitude estimate for some of the
parameters entering the matrix (18), by comparing 
the evolution dictated by Eq.(4) with experimentally
measured quantities. 

Any physical property of the $K$-$\overline K$ system can
be extracted from the density matrix $\rho(t)$ by taking
its trace with suitable hermitian operators. Useful
observables are associated with the decays of the neutral
kaons into $2\pi$ or $3\pi$ states, or into semileptonic
states $\pi\ell\nu$.

We take the approximation in which only 
the $K$ meson can decay into the state $\pi^-\ell^+\nu$ (therefore ignoring
small violations of the so called $\Delta S=\Delta Q$ rule [15]).
In the basis $K_1$, $K_2$ of Eq.(10), the operator responsible for the decay
in this final state is proportional to:
$$
{\cal O}_{\ell^+}=\left(\matrix{1&1\cr
                                1&1\cr}\right)\ .\eqno(21)
$$
Similarly, for the decay into the state $\pi^+\ell^-\nu$ one has:
$$
{\cal O}_{\ell^-}=\left(\matrix{\phantom{-}1&-1\cr
                                -1&\phantom{-}1\cr}\right)\ .\eqno(22)
$$
With these operators, one can form an experimentally testable observable,
the so called $CP$ violating charge asymmetry [9, 14]:
$$
\delta(t)={
{\rm Tr}\Big[\rho(t)\left({\cal O}_{\ell^+}-{\cal O}_{\ell^-}\right)\Big]
\over
{\rm Tr}\Big[\rho(t)\left({\cal O}_{\ell^+}+{\cal O}_{\ell^-}\right)\Big]}
\ .\eqno(23)
$$

The decay of neutral kaons into pions are described by the two
operators [14, 16]:
$$
{\cal O}_{2\pi}=\left(\matrix{1&0\cr
                              0&0\cr}\right)\ ,\qquad
{\cal O}_{3\pi}=\left(\matrix{0&0\cr
                              0&1\cr}\right)\ .  \eqno(24)
$$
Out of these operators one can form
two new observables, the decay rates of the neutral kaon system
into $2\pi$ and $3\pi$:
$$
R_{2\pi}(t)={
{\rm Tr}\Big[\rho(t){\cal O}_{2\pi}\Big]\over
{\rm Tr}\Big[\rho(0){\cal O}_{2\pi}\Big]}\ ,\qquad
R_{3\pi}(t)={
{\rm Tr}\Big[\rho(t){\cal O}_{3\pi}\Big]\over
{\rm Tr}\Big[\rho(0){\cal O}_{3\pi}\Big]}\ .\eqno(25)
$$

To compute the observables $\delta(t)$ and $R(t)$, one needs to solve
the evolution equations (4) and (8) for the density matrix (11).
We shall explicitly give only the asymptotic expression for the 
solutions of those equations in the long and short time regime. 
A more detailed and complete treatment,
which allows a comparison of the quantities (23) and (25) with
the experimental data also at intermediate times, will be reported elsewhere.

One can check that for large times, the solution of Eq.(4) with $h[\rho]$ as in
(8), decays exponentially to
$$
\rho_L=\left(\matrix{
|\epsilon_L|^2-{h_{33}\over 2\Delta{\mit\Gamma}}-
\left|{h_{23}+i h_{13}\over 2\Delta\lambda}\right|^2
-2{\Delta m\over\Delta{\mit\Gamma}}\, {\cal I}m\left[\epsilon_L 
{h_{23}-i h_{13}\over(\Delta\lambda)^*}\right] &
\epsilon_L-{h_{23}+i h_{13}\over 2\Delta\lambda}\cr
\epsilon_L^*-{h_{23}-i h_{13}\over 2(\Delta\lambda)^*} & 1}\right)\ ,\eqno(26)
$$
while for short times, the behavior of $\rho(t)$ is again exponential
but with different coefficients:
$$
\rho_S=\left(\matrix{ 1 & 
\epsilon_S^*+{h_{23}+i h_{13}\over 2(\Delta\lambda)^*}\cr
\epsilon_S+{h_{23}-i h_{13}\over 2\Delta\lambda} &
|\epsilon_S|^2+{h_{33}\over 2\Delta{\mit\Gamma}}-
\left|{h_{23}-i h_{13}\over 2\Delta\lambda}\right|^2
+2{\Delta m\over\Delta{\mit\Gamma}}\, {\cal I}m\left[\epsilon_S 
{h_{23}+i h_{13}\over(\Delta\lambda)^*}\right]}\right)\ ;\eqno(27)
$$
both these asymptotic solutions are independent from the initial conditions.
In the previous formulas, $\Delta\lambda=\lambda_L-\lambda_S=
\Delta m+i\Delta{\mit\Gamma}/2$
denotes the difference of the eigenvalues (13) of
the Weisskopf-Wigner hamiltonian, with $\Delta m=m_L-m_S$ the mass difference
of the $K_L$ and $K_S$ particles, and $\Delta{\mit\Gamma}=\gamma_S-\gamma_L$ 
the
difference of their decay widths. The two density matrices
$\rho_L$ and $\rho_S$ describe mixed states, corresponding to a mixture
of the conventional $K_L$ and $K_S$ neutral kaons. 

On general grounds [9, 14, 16], one expects the values of the parameters
$h_{rs}$ to be very small. For example, 
assuming that the extra term $h[\rho(t)]$
in the evolution equation (4) is due to the effects of quantum gravity,
a rough estimate for $h_{rs}$ is $m_L^2/M_P\sim 10^{-20}\ {\rm GeV}$, 
where $M_P$ is the Planck mass. In writing $\rho_L$ and $\rho_S$
above, we have taken into account this fact and 
kept only the leading contributions in each entry of the two matrices.
This approximation will suffice for the comparison with the experimental
data to the accuracy we require.

Inserting (26) and (27) into the expressions for the observables $\delta(t)$
and $R(t)$, one easily gets, to lowest order:
$$
\delta_L=2\, {\cal R}e\left[\epsilon_L-{h_{23}
+i h_{13}\over 2\Delta\lambda}\right]\ ,\quad
\delta_S=2\, {\cal R}e\left[\epsilon_S+{h_{23}-
i h_{13}\over 2\Delta\lambda}\right]\ ,\eqno(28)
$$
and
$$\eqalign{&R^L_{2\pi}=|\epsilon_L|^2-{h_{33}\over 2\Delta{\mit\Gamma}}-
\left|{h_{23}+i h_{13}\over 2\Delta\lambda}\right|^2
-2{\Delta m\over\Delta{\mit\Gamma}}\ {\cal I}m\left[\epsilon_L 
{h_{23}-i h_{13}\over(\Delta\lambda)^*}\right]\ ,\cr
           &R^S_{3\pi}=|\epsilon_S|^2+{h_{33}\over 2\Delta{\mit\Gamma}}-
\left|{h_{23}-i h_{13}\over 2\Delta\lambda}\right|^2
+2{\Delta m\over\Delta{\mit\Gamma}}\ {\cal I}m\left[\epsilon_S 
{h_{23}+i h_{13}\over(\Delta\lambda)^*}\right]\ .}\eqno(29)
$$
One can now express the parameters $h_{23}$, $h_{13}$ and $h_{33}$
in terms of these phenomenological quantities. Ignoring $CPT$-violating
effects in the Weisskopf-Wigner hamiltonian, $\epsilon_L=\epsilon_S=\epsilon$,
and noticing that $\Delta m\simeq \Delta{\mit\Gamma}/2$, one derives
from (28) and (29):
$$
\eqalignno{&h_{23}\simeq 2(\delta_S-\delta_L)\, \Delta m\ , &(30a)\cr
          &h_{13}\simeq\left[4\, {\cal R}e(\epsilon)-(\delta_L+\delta_S)
\right] \Delta m\ , &(30b)\cr
          &h_{33}\simeq\left(R^S_{2\pi}-R^L_{3\pi}\right) \Delta m
-2\, h_{23}\, {\cal R}e(\epsilon)\ . &(30c)}
$$
Although not all the phenomenological parameters appearing in the r.h.s. are 
well measured, from the existing experimental data [17] the following order
of magnitude upper bounds can be deduced:
$$ 
|h_{23}|\leq 10^{-17}\, {\rm GeV}\ ,\quad
|h_{13}|\leq 10^{-18}\, {\rm GeV}\ ,\quad
|h_{33}|\leq 10^{-20}\, {\rm GeV}\ . \eqno(31)
$$

To obtain an estimate on the values of the remaining 
three parameters in the matrix $[h_{rs}]$ one needs a more
detailed treatment based on solutions of the evolution
equations Eq.(4) and Eq.(8) valid also at intermediate times
and not only in the asymptotic long- and short-time regions;
work along these lines is in progress.
Nevertheless, notice that inequalities (19) give some partial constraints
on these parameters: $|h_{12}|\leq |h_{33}|/2$ and 
$|h_{11}-h_{22}|\leq |h_{33}|$.
Since $|h_{33}|$ is very small, one expects $|h_{12}|$ to be even smaller and
$|h_{11}|$ of the same order of magnitude of $|h_{22}|$.


\bigskip
\centerline{\bf Acknoledgement}
\medskip

We thank N. Paver for many useful discussions and comments.


\vfill\eject

\centerline{\bf References}
\medskip

\item{1.} {\it The Second Da$\phi$ne Physics Handbook}, L. Maiani, G. Pancheri
and N. Paver, eds., (INFN, Frascati, 1995)
\smallskip
\item{2.} S. Hawking, Phys. Rev. D {\bf 14} (1975) 2460; Comm. Math. Phys.
{\bf 43} (1975) 199
\smallskip
\item{3.} S. Hawking, Comm. Math. Phys. {\bf 87} (1983) 395
\smallskip
\item{4.} V. Gorini, A. Frigerio, M. Verri, A. Kossakowski and
E.C.G. Surdarshan, Rep. Math. Phys. {\bf 13} (1978) 149
\smallskip
\item{5.} G. Lindblad, Comm. Math. Phys. {\bf 48} (1976) 119
\smallskip
\item{6.} E.B. Davies, {\it Quantum Theory of Open Systems}, (Academic Press,
New York, 1976)
\smallskip
\item{7.} K. Kraus, {\it States, Effects, and Operations}, Lect. Notes Phys.
{\bf 190} (Springer Verlag, Berlin, 1983)
\smallskip
\item{8.} F. Benatti and H. Narnhofer, Lett. Math. Phys. {\bf 15} (1988) 325
\smallskip
\item{9.} J. Ellis, J.S. Hagelin, D.V. Nanopoulos and M. Srednicki,
Nucl Phys. {\bf B241} (1984) 381
\smallskip
\item{10.} L. Fonda, G.C. Ghirardi and A. Rimini, Rep. Prog. Phys.
{\bf 41} (1978) 587 
\smallskip
\item{11.} T. Banks, L. Susskind and M.E. Peskin, Nucl. Phys. {\bf B244}
(1984) 125
\smallskip
\item{12.} M. Srednicki, Nucl. Phys. {\bf B410} (1933) 143
\smallskip
\item{13.} W. Unruh and R.M. Wald, Phys. Rev. D {\bf 52} (1995) 2176
\smallskip
\item{14.} P. Huet and M.E. Peskin, Nucl. Phys. {\bf B434} (1995) 3
\smallskip
\item{15.} For example, see L. Maiani in [1]
\smallskip
\item{16.} J. Ellis, J.L. Lopez, N.E. Mavromatos and D.V. Nanopoulos,
Precision tests of $CPT$ symmetry and quantum mechanics in the neutral kaon
system, CERN-TH.95-99 ({\tt hep-ph/9505340}) 
\smallskip
\item{17.} Particle Data Group, Phys. Rev. D {\bf 50} (1994) 1173;
R. Adler {\it et al.}, Nucl. Phys. {\bf A558} (1993) 449c;
R. Adler {\it et al.}, Serach for $CP$ violation in the decay of neutral
kaons to $\pi^+\pi^-\pi^0$, CERN-PPE/95-189

\bye